\documentclass[pre,showpacs,preprint]{revtex4}
\usepackage{amsmath}

\begin{document}
\title{First order phase transitions in nanoscopic systems}
\author{A. Boer}
\date{\today}
\email{boera@unitbv.ro}
\affiliation{Department of Physics, Transilvania University, B-dul
Eroilor, R-2200, Bra\c {s}ov, Romania}

\author{S. Dumitru}
\email{s.dumitru@unitbv.ro}
\affiliation{Department of Physics, Transilvania University, B-dul
Eroilor, R-2200, Bra\c {s}ov, Romania}

\begin{abstract}
The problem of first order phase transitions in nanoscopic systems is investigated in the framework of Hill's nanothermodynamics. We obtain the equilibrium conditions and a generalized version of the Clapeyron-Clausius equation for a nanoscopic system which contains two phases. Our study is exemplified for the case when one of the phases consists of an ideal gas.
\end{abstract}

\maketitle

\section{Introduction}\label{sec:1}
In the last decade were published many theoretical articles \cite{Hill1, Hill2, Rajagopal, Hill3,
Carrete} regarding the thermodynamical or statistical study of nanoscopic systems. We can mention in this context Hill's nanothermodynamics, quantum thermodynamics and some applications of nonextensive thermostatistics in the mesoscopic domain. In the present paper, by taking into account Hill's theory we present an investigation of the first order phase transitions in nanoscopic systems. We will establish the equilibrium conditions for a nanoscopic system which contains two phases and the generalization of the Clapeyron-Clausius equation for nanoscopic systems.

\section{Theory}\label{sec:2}
To establish the equilibrium conditions in the case of nanoscopic systems we will use Hill's approach, i.e. we consider an ensemble of $\mathcal{N}$ identical nanoscopic systems which represents a macroscopic system \cite{Hill1}.

Let us consider a macroscopic system which contains two phases in thermodynamical equilibrium. Suppose that we divide each phase in identical nanoscopic systems. The differential of internal energy for the considered phases is given by \cite{Hill1}
\begin{equation}\label{eq:1}
dE_{1t} = T_1 \, dS_{1t} - p_1 \, dV_{1t} + \mu_1 \, dN_{1t} + \mathcal{E}_1\, d\mathcal{N}_1
\end{equation}
\begin{equation}\label{eq:2}
dE_{2t} = T_2 \, dS_{2t} - p_2 \, dV_{2t} + \mu_2 \, dN_{2t} + \mathcal{E}_2\, d\mathcal{N}_2
\end{equation}
In the above relations we have used the standard notations, i.e. $T$-temperature, $S$-entropy, $p$-pressure, $V$-volume, $\mu$-chemical potential and $N$-particle number. $\mathcal{E}$ represents the generalized thermodynamical potential introduced by Hill and $\mathcal{N}$ signifies the number of nanoscopic systems in the ensemble. The indexes 1 and 2 refer to the considered phases while the index 
``$t$'' signifies the fact that the respective quantity regards the ``total'' ($t$) ensemble of identical nanoscopic systems.

The thermodynamical equilibrium state is given by the maximum of the entropy, which lead to the following extremum condition
\begin{equation}\label{eq:3}
\delta S_t = \delta S_{1t} + \delta S_{2t} = 0
\end{equation}
Taking into account the relations \eqref{eq:1} and \eqref{eq:2} one obtains
\begin{equation}\label{eq:4}
\frac{1}{T_1} \delta E_{1t} + \frac{1}{T_2} \delta E_{2t} + \frac{p_1}{T_1} \delta V_{1t} + 
\frac{p_2}{T_2} \delta V_{2t} - \frac{\mu_1}{T_1} \delta N_{1t} - \frac{\mu_2}{T_2} \delta N_{2t}
- \frac{\mathcal{E}_1}{T_1} \delta{\mathcal{N}_1} - \frac{\mathcal{E}_2}{T_2} \delta{\mathcal{N}_2}
= 0
\end{equation}   
As the global system is isolated, one finds the conditions:
\begin{equation}\label{eq:5}
\begin{matrix}
E_t = E_{1t} + E_{2t} = \text{const.} \, ; & \quad \delta E_{2t} = - \delta E_{1t} \\
V_t = V_{1t} + V_{2t} = \text{const.} \, ; & \quad \delta V_{2t} = - \delta V_{1t} \\
N_t = N_{1t} + N_{2t} = \text{const.} \, ; & \quad \delta N_{2t} = - \delta N_{1t} \\
\mathcal{N} = \mathcal{N}_1 + \mathcal{N}_2 = \text{const.} \, ; & \quad
\delta\mathcal{N}_2 = - \delta\mathcal{N}_1
\end{matrix}
\end{equation}
Then the relation \eqref{eq:4} takes the following form
\begin{equation}\label{eq:6}
\left( \frac{1}{T_1} - \frac{1}{T_2} \right) \, \delta E_{1t} + \left( \frac{p_1}{T_1} -
\frac{p_2}{T_2} \right) \, \delta V_{1t} - \left( \frac{\mu_1}{T_1} - \frac{\mu_2}{T_2} \right) 
\, \delta N_{1t} - \left( \frac{\mathcal{E}_1}{T_1} - \frac{\mathcal{E}_2}{T_2} \right) \,
\delta \mathcal{N}_1 = 0
\end{equation}
The above relation can be satisfied for arbitrary variations of $E_{1t}$, $V_{1t}$, $N_{1t}$ and $\mathcal{N}_1$ only if
\begin{equation}\label{eq:7}
T_1 = T_2 = T
\end{equation}
\begin{equation}\label{eq:8}
p_1 = p_2 = p
\end{equation}
\begin{equation}\label{eq:9}
\mu_1 = \mu_2 = \mu
\end{equation}
\begin{equation}\label{eq:10}
\mathcal{E}_1 (T, p, \mu) = \mathcal{E}_2 (T, p, \mu)
\end{equation}
The conditions \eqref{eq:7}-\eqref{eq:9} are well known from macroscopic thermodynamics. But we must mention that in the case of nanoscopic systems $T$, $p$ and $\mu$ are independent variables \cite{Hill3,Rajagopal}. We observe also that for nanoscopic systems we have a supplementary condition, given by the equality of Hill's potentials $\mathcal{E}_1$ and $\mathcal{E}_2$ for the considered phases. In the case of macroscopic systems the condition \eqref{eq:10} become the trivial equality $0=0$. 

Because $E_{1t}$ is a first order homogeneous function with respect to $S_{1t}$, $V_{1t}$, $N_{1t}$ and $\mathcal{N}_1$, we have \cite{Hill3}
\begin{equation}\label{eq:11}
E_{1t} = T_1  S_{1t} - p_1 V_{1t} + \mu_1 N_{1t} + \mathcal{E}_1 \mathcal{N}_1
\end{equation}
Differentiating the above equation and taking into account the relation \eqref{eq:1} one obtains
\begin{equation}\label{eq:12}
\mathcal{N}_1 \, d\mathcal{E}_1 = - S_{1t}\, dT_1 + V_{1t}\, dp_1 - N_{1t}\, d\mu_1 
\end{equation}
On the other hand $S_{1t} = S_1 \mathcal{N}_1$, $V_{1t} = V_1 \mathcal{N}_1$, 
$N_{1t} = N_1 \mathcal{N}_1$. So we obtain the following equation \cite{Hill3}
\begin{equation}\label{eq:13}
d\mathcal{E}_1 = -S_1 \, dT_1 + V_1 \, dp_1 - N_1 \, d\mu_1
\end{equation}
where the quantities $S_1$, $V_1$ and $N_1$ refer to a nanoscopic system.
The above equation represents a generalization of the Gibbs-Duhem equation from macroscopic thermodynamics.

In a similar way we obtain
\begin{equation}\label{eq:14}
d\mathcal{E}_2 = -S_2 \, dT_2 + V_2 \, dp_2 - N_2 \, d\mu_2
\end{equation}
The equilibrium conditions \eqref{eq:7}-\eqref{eq:10} with the relations \eqref{eq:13} and 
\eqref{eq:14} lead to the following equation
\begin{equation}\label{eq:15}
-S_1 \, dT_1 + V_1 \, dp_1 - N_1 \, d\mu_1 = -S_2 \, dT_2 + V_2 \, dp_2 - N_2 \, d\mu_2
\end{equation}

In the case of a nanoscopic system the quantities $T$, $p$ and $\mu$ are independent variables and the chemical potential $\mu$ depends on $T$, $p$ and $N$ \cite{Hill3,Rajagopal}. Note that in the case of macroscopic systems the chemical potential depends only on $T$ and $p$ \cite{Rumer}.

Let us focus our attention on the case of nanoscopic systems, for which we have
\begin{equation*}
d\mu = \left( \frac{\partial\mu}{\partial T} \right)_{p,N_2} dT +
\left( \frac{\partial\mu}{\partial p} \right)_{T,N_2} dp +
\left( \frac{\partial\mu}{\partial N_2} \right)_{T,p} dN_2
\end{equation*}
Introducing the expression of $d\mu$ in relation \eqref{eq:15} one obtains
\begin{eqnarray}\label{eq:16}
S_2 - S_1 + \left( N_2 - N_1 \right) \left( \frac{\partial\mu}{\partial T}\right)_{p,N_2}
- \left[ V_2 - V_1 - \left( N_2 - N_1 \right) \left( \frac{\partial\mu}{\partial p}\right)_{T,N_2}
\right] \frac{dp}{dT} + \nonumber \\
+ \left( N_2 - N_1 \right) \left( \frac{\partial\mu}{\partial N_2}\right)_{T,p} \frac{dN_2}{dT} = 0
\end{eqnarray}
This equation represents a generalization of the well known Clapeyron-Clausius equation from macroscopic thermodynamics.

In the case of macroscopic systems $\mu$ depends only on $T$ and $p$, therefore we have
\begin{equation}\label{eq:17}
\frac{dp}{dT} = \frac{ \frac{N_2}{N_1} s_2 - s_1 + \left( \frac{N_2}{N_1} - 1 \right)
\left( \frac{\partial\mu}{\partial T}\right)_{p}  }
{\frac{N_2}{N_1} v_2 - v_1 - \left( \frac{N_2}{N_1} - 1 \right)
\left( \frac{\partial\mu}{\partial p}\right)_{T} }
\end{equation}
where $s = \frac{S}{N}$ and $ v = \frac{V}{N}$. In the limit $N_1 \to \infty$, $N_2 \to \infty$
we obtain the usual Clausius-Clapeyron equation \cite{Rumer}
\begin{equation}\label{eq:18}
\frac{dp}{dT} = \frac{s_2 - s_1}{v_2 - v_1}
\end{equation}

\section{The analysis of a particular case}\label{sec:3}
In this section we will present briefly an application of the theoretical results obtained above for the case when one of the phases (denoted as phase 2) consists of an ideal gas. The chemical potential has the form \cite{Hill3}
\begin{equation}\label{eq:19}
\mu_2 = kT \ln \left( \frac{N_2}{1+ N_2} \frac{p\Lambda^3}{kT} \right)
\end{equation}
where $\Lambda = h / \left( 2\pi mkT \right)^{1/2}$, $h$ being Planck's constant, $k$ Boltzmann's constant and $m$ the mass of a molecule. Then one obtains
\begin{equation}\label{eq:20}
\left( \frac{\partial \mu_2}{\partial T} \right)_{p,N_2} = k \left[ \ln \left(
\frac{N_2}{1+N_2} \frac{p\Lambda^3}{kT} \right) - 1 \right]
\end{equation}
\begin{equation}\label{eq:21}
\left( \frac{\partial \mu_2}{\partial p} \right)_{T,N_2} = \frac{kT}{p}
\end{equation}
\begin{equation}\label{eq:22}
\left( \frac{\partial \mu_2}{\partial N_2} \right)_{T,p} = \frac{kT}{N_2 ( 1 + N_2) }
\end{equation}
Equation \eqref{eq:16} gives
\begin{eqnarray}
\left( S_2 - S_1 \right) + k \left( N_2 - N_1 \right) \left[ \ln \left( \frac{N_2}{1+N_2} \frac{p\Lambda^3}{kT} \right) -1 \right] 
- \left[ V_2 - V_1 - \left( N_2 - N_1 \right) \frac{kT}{p} \right] \frac{dp}{dT} + \nonumber \\
 + \left( N_2 - N_1 \right) \frac{kT}{N_2 \left( 1 + N_2 \right) } \frac{dN_2}{dT} = 0
\end{eqnarray}
This is the generalized Clapeyron-Clausius formula for the case of a nanoscopic system which contains two phases, one of them being an ideal gas.

\section{Conclusions}\label{sec:4}
In the present work we have studied the problem of first order phase transitions in nanoscopic systems from a thermodynamical point of view. We established the equilibrium conditions for a nanoscopic system which contains two phases, based on Hill's theory. The fact that in the case of nanoscopic systems the quantities $T$, $p$ and $\mu$ are independent variables has some interesting consequences, one of them being a supplementary equilibrium condition, in addition to the ones known from macroscopic thermodynamics. This condition is connected with the generalized potential $\mathcal{E}$ introduced by Hill. We obtained also a generalized form of the Clapeyron-Clausius equation and we studied briefly a particular case corresponding to the situation when one of the phases contains an ideal gas.




\begin{thebibliography}{99}
\bibitem{Hill1} T.L. Hill, Nano Lett. Vol.1, No.5, 273-275 (2001).
\bibitem{Hill2} T.L. Hill, Nano Lett. Vol.1, No.3, 159-160 (2001).
\bibitem{Rajagopal} A. K. Rajagopal , C. S. Pande , Sumiyoshi Abe, arXiv:cond-mat/0403738 (2004).
\bibitem{Hill3} T.L. Hill, Nano Lett. Vol.2, No.6, 609-613 (2002).
\bibitem{Carrete} J. Carrete, L. M. Varela, and L. J. Gallego, Phys. Rev. E \textbf{77}, 022102 (2008).
\bibitem{Rumer} Yu. B. Rumer, M. Sh. Ryvkin, \textit{Thermodynamics, Statistical Physics and Kinetics}, Mir Publishers, Moscow 1980.
\end{thebibliography}
\end{document}